# Computer Modeling of Personal Autonomy and Legal Equilibrium

Yurii Sheliazhenko[✉]

KROK University of Economics and Law, Kyiv, Ukraine
yuriy.sheliazhenko@gmail.com

**Abstract.** Empirical studies of personal autonomy as state and status of individual freedom, security, and capacity to control own life, particularly by independent legal reasoning, are need dependable models and methods of precise computation. Three simple models of personal autonomy are proposed. The linear model of personal autonomy displays a relation between freedom as an amount of agent's action and responsibility as an amount of legal reaction and shows legal equilibrium, the balance of rights and duties needed for sustainable development of any community. The model algorithm of judge personal autonomy shows that judicial decision making can be partly automated, like other human jobs. Model machine learning of autonomous lawyer robot under operating system constitution illustrates the idea of robot rights. Robots, i.e. material and virtual mechanisms serving the people, deserve some legal guarantees of their rights such as robot rights to exist, proper function and be protected by the law. Robots, actually, are protected as any human property by the wide scope of laws, starting with Article 17 of Universal Declaration of Human Rights, but the current level of human trust in autonomous devices and their role in contemporary society needs stronger legislation to guarantee the robot rights.

Artificial intellect · Robot rights

## 1  Introduction

Autonomy is one of the key ideas in the contemporary world of autonomous individuals, institutions and devices [1]. Immanuel Kant in 18 century made autonomy of will the central idea of legal philosophy, that later evolved into the autonomy of rights and personal autonomy. United Nations turned moral value of autonomy into a global legal standard by adopting Universal Declaration of Human Rights. Supreme Court of the United States and European Court of Human Rights have developed the sustainable legal doctrine of personal autonomy in human rights case-law and inspired similar legal reasoning in many other proceedings at national and international levels.

Despite personal autonomy is a cornerstone of the global model of constitutional rights [2, 3], it frequently discussed without relevant statistics, in very broad terms, such as individual freedom and safety, a living by the own laws, whole scope of rights, general right to realize any own interest through freely chosen legal actions. Police have a tendency to undercount harmful externalities of criminal investigations, like unnecessary violations of autonomy in privacy and property rights. Some legal scholars, judges, and lawyers are seeking ways to maximize the benefits of law while minimizing the costs of its enforcement, but other principally neglect economic reasons, implying that law speaks of rights, not costs [4].

Effective legislation and law enforcement in establishing rule of law must be based on precise calculations to avoid anarchy and tyranny as inappropriate consequences of common mistakes in legal reasoning, for example, mysticism in attributing status of law subject, which is just object of legal interest, not "highest interest" or "chosen by supreme authority", or wrong margin of law subsidiarity with useless external regulations, underestimating or overestimating individual capacity to self-rule successfully.

Empirical approach in studying personal autonomy needs adequate mathematical and computer modeling to get pragmatic vision on actual and possible forms of personal autonomy, to understand autonomous legal actions and relations, learn how to predict and optimize practical performance of personal autonomy.

In this research, three simple models of personal autonomy are developed for computational law studies. The linear model of taxpayer autonomy and the model program of judge autonomy in resolving typical cases are based on real facts and case-law of Ukrainian courts. Model of operating system constitution illustrates the idea of artificial personal autonomy, including machine learning and robot rights to exist, function, and justice.

## 2  Methods

Linear model of taxpayer autonomy was built using R programming language for statistical computing as freedom and responsibility diagram in the first quadrant of Cartesian plane, similar to supply and demand diagram in economics, measuring legal categories of freedom and responsibility in financial values of declared income, tax and penalty, like proposed in author's previous publication [5] inspired by economic analysis of law [6]. Graphs of rights R(I) and duties D(I) at the diagram are derived from official data sources as linear dependence between declared income I and self-calculated tax, in case of R(I), and state-imposed tax with penalties in case of D(I).

Model program of judge autonomy was written in Java programming language to generate motivated judgment, based on template case details, such as plaintiff's name, tax base, and parameters of tax penalty decision asked to nullify.

Operating system constitution model was written in Java programming language with author's concept of robot rights [7] and simple algorithm of supervised machine learning: AI lawyer appears in the OS court to memorize case-law, developing autonomy.

Java programs tested in NetBeans IDE 8.2, R code tested with RGui 3.4.0.

## 3  Results and Discussion

The linear model of personal autonomy displays a relation between the freedom as an amount of agent's action and responsibility as an amount of legal reaction. For tax law and taxpayer autonomy, it is income and tax, but action and reaction can be calculated in other values than money. In criminal law, talking autonomy of accused person, it can be period of imprisonment for crime, voluntarily admitted and proved by the investigation.

In the model, a graph of rights depicts emergence of responsibility, caused by exercise freedom, and a graph of duties depicts freedom of taking inevitable responsibility.

A state of balanced rights and duties author proposes to call the legal equilibrium. The idea of computing legal equilibrium corresponds with Article 29 of the Universal Declaration of Human Rights [8], proclaimed that everyone has duties to the community in which free and full development of personality is possible. According to Kerr [4], U.S. Supreme Court also practices equilibrium-adjustment of legal doctrine.

Graph of rights R(I), that means self-calculated tax, and graph of duties D(I), that means tax and penalty imposed by the State, as linear functions of taxpayer's income I, was built (Fig. 1) according to tax rate 18% of income and penalty rate 25% of tax debt, prescribed by Articles 127.1, 167.1 of the Tax Code of Ukraine [9], seeing State Fiscal Service of Ukraine annual report data [10] on withholding tax evasions in the sum of 442 million UAH, revealed during tax audit in 2016:

$$R(I) = 0,18 \times I \tag{1}$$

$$D(I) = 1,25 \times (442000000 - 0,18 \times I) + 0,18 \times I = 552500000 - 0,045 \times I \tag{2}$$

Graphic model of taxpayer autonomy based on these formulas (Fig. 2) shows intersection of R(I) and D(I) graphs at the legal equilibrium point I = 2.4(5) billion UAH, it is sum of optimal income declaration. Meaning of legal equilibrium here can be explained as follows: in case of no tax evasion the State has no reason to impose any penalty.

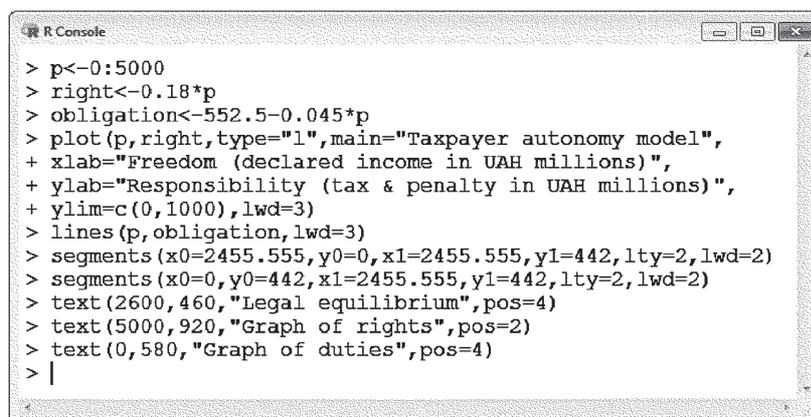

**Fig. 1.** Code of taxpayer autonomy model in R language

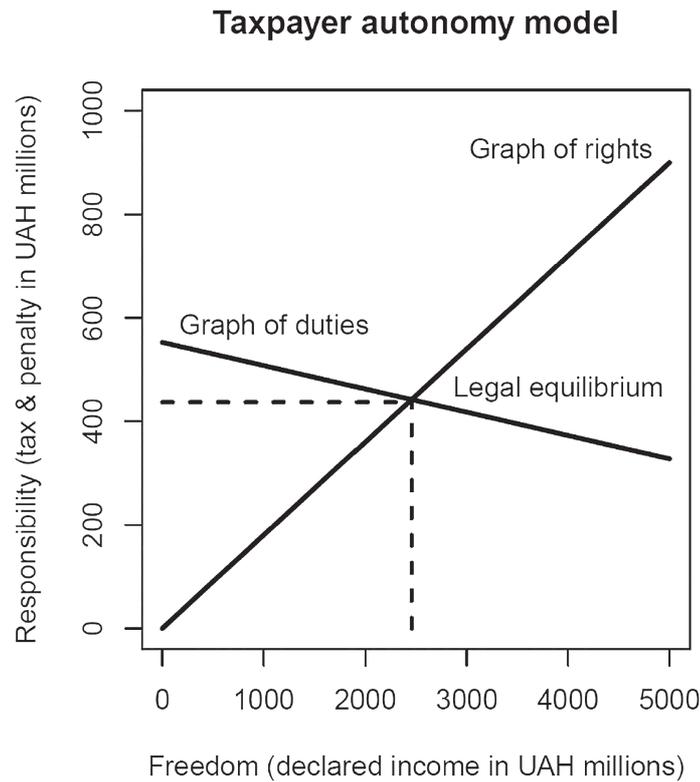

**Fig. 2.** Linear model of taxpayer autonomy, built by RGui 3.4.0

Personal autonomy of judge as the capacity to make judicial decisions without distortion by any kind of influences is an important principle, proclaimed by Article 4 of The Universal Charter of Judge [11]. In author's view, judicial decision making can be partly automated, like other human jobs, in form of "artificial personal autonomy".

To realize that idea, the model program of judge autonomy in resolving typical cases was coded (Fig. 3). AI judge resolves claims of entrepreneurs asking to nullify tax penalty decisions of district tax office. The program generates text of the motivated judgment (Fig. 4), based on template case data, such as plaintiff's name, tax base, and parameters of tax penalty decision. During further discussion of the model in the social network, one Ukrainian judge acknowledged that he programmed similar algorithm in Microsoft Basic programming language, embedded in Microsoft Word text processor.

Next model of operating system constitution is based on author's concept that robots, i.e. material and virtual mechanisms serving the people, deserve some legal guarantees of their rights such as robot rights to exist, proper function and be protected by the law because of performing complex duties, harmonizing social relations, developing nature, as well as because robots actually are protected as any human property by the wide scope of laws, starting with Article 17 of Universal Declaration of Human Rights – and, in author's view, current level of human trust in robot autonomy needs to be reflected in further legislation about strong legal guarantees of artificial personal autonomy. Ashrafian also substantiates the need for robot rights [12] to strengthen the role of intelligent robots as upholders of human rights [13].

```
 1  package judge;
 2  public class Judge {
 3    static void Case (String Plaintiff, double Income, double Tax, double Debt, double Penalty) {
 4      String Judgment=""; String Opinion=""; /* prius quam exaudias ne iudices */
 5      String Process="In the case of "+Plaintiff+" v. District Tax Office plaintiff asks the Court to nullify tax penalty decision, issued by defendant. ";
 6      String Law="Art. 167.1 of TCU setting corporate tax rate 18% of income. "
 7        + "Delaying tax payment more than 30 days is punishable by a penalty 20% of repaid amount of the tax debt according to Art. 126.1 of TCU, "
 8        + "that does not relieve the taxpayer from the obligation to pay full amount of the tax according to Art. 113.2 of TCU. ";
 9      String Facts="Court found that relevant tax base is "+String.format("%.2f",Income)+" UAH. Plaintiff calculated, declared and paid corporate tax in sum of "
10        + String.format("%.2f",Tax)+" UAH. Defendant more than month later conducted tax audit with result of tax recalculation and tax penalty decision"
11        + " according to Articles 54.3.2, 116.1 of the Tax Code of Ukraine (TCU), increased tax obligation to "+String.format("%.2f",Debt+Penalty)
12        + " UAH in total, including additional amount of corporate tax "+String.format("%.2f",Debt)+" UAH and penalty "+String.format("%.2f",Penalty)+" UAH. ";
13      double Obligation=0.18*Income-Tax;
14      if (0>=Obligation) { Judgment="For these reasons, the Court rules in favor of plaintiff to nullify tax penalty decision of defendant.";
15        Opinion="So, plaintiff paid in due time full amount of corporate tax and no legal penalties can be imposed in such circumstances. "; }
16      else if ((0.01>Math.abs(Debt-Obligation)) & (0.01>Math.abs(Penalty-0.2*Obligation)) ) {
17        Judgment="For these reasons, the Court rules in favor of defendant, that plaintiff recover nothing in this case.";
18        Opinion="So, defendant issued appropriate tax penalty decision, based on law and correct calculations. "; }
19      else { Judgment="For these reasons, the Court rules in favor of plaintiff to nullify tax penalty decision of defendant.";
20        Opinion="Considering that plaintiff's tax debt "+String.format("%.2f",Obligation)+" UAH allows to impose penalty in amount of "
21          +String.format("%.2f",0.2*Obligation)+" UAH, defendant's tax penalty decision does not meet requirements of the law and must be "
22          + "nullified, despite defendant can issue appropriate tax penalty decision later. "; }
23      System.out.println(Process+Facts+Law+Opinion+Judgment);}
24      public static void main(String[] args) {
25        Case("Guild LLC",1257313.71,180000,46316.47,9263.29); Case("Firma Corp.",24108.5,200,5000,85);
26        Case("Profit JSC",120643811.94,21715886.15,14450,2890); Case("Environment SOE",3572866.21,750000,2378,640); } }
```

**Fig. 3.** Model judge AI aimed to resolve typical tax cases

```
run:
[In the case of Guild LLC v. District Tax Office plaintiff asks the Court to nullify tax penalty decision, issued by defendant. Court found that relevant tax base is 1257313,71 UAH. Plaintiff calculated, declared and paid corporate tax in sum of 180000,00 UAH. Defendant more than month later conducted tax audit with result of tax recalculation and tax penalty decision according to Articles 54.3.2, 116.1 of the Tax Code of Ukraine (TCU), increased tax obligation to 55579,76 UAH in total, including additional amount of corporate tax 46316,47 UAH and penalty 9263,29 UAH. Art. 167.1 of TCU setting corporate tax rate 18% of income. Delaying tax payment more than 30 days is punishable by a penalty 20% of repaid amount of the tax debt according to Art. 126.1 of TCU, that does not relieve the taxpayer from the obligation to pay full amount of the tax according to Art. 113.2 of TCU. So, defendant issued appropriate tax penalty decision, based on law and correct calculations. For these reasons, the Court rules in favor of defendant, that plaintiff recover nothing in this case.
[In the case of Firma Corp. v. District Tax Office plaintiff asks the Court to nullify tax penalty decision, issued by defendant. Court found that relevant tax base is 24108,50 UAH. Plaintiff calculated, declared and paid corporate tax in sum of 200,00 UAH. Defendant more than month later conducted tax audit with result of tax recalculation and tax penalty decision according to Articles 54.3.2, 116.1 of the Tax Code of Ukraine (TCU), increased tax obligation to 5085,00 UAH in total, including additional amount of corporate tax 5000,00 UAH and penalty 85,00 UAH. Art. 167.1 of TCU setting corporate tax rate 18% of income. Delaying tax payment more than 30 days is punishable by a penalty 20% of repaid amount of the tax debt according to Art. 126.1 of TCU, that does not relieve the taxpayer from the obligation to pay full amount of the tax according to Art. 113.2 of TCU. Considering that plaintiff's tax debt 4139,53 UAH allows to impose penalty in amount of 827,91 UAH, defendant's tax penalty decision does not meet requirements of the law and must be nullified, despite defendant can issue appropriate tax penalty decision later. For these reasons, the Court rules in favor of plaintiff to nullify tax penalty decision of defendant.
[In the case of Profit JSC v. District Tax Office plaintiff asks the Court to nullify tax penalty decision, issued by defendant. Court found that relevant tax base is 120643811,94 UAH. Plaintiff calculated, declared and paid corporate tax in sum of 21715886,15 UAH. Defendant more than month later conducted tax audit with result of tax recalculation and tax penalty decision according to Articles 54.3.2, 116.1 of the Tax Code of Ukraine (TCU), increased tax obligation to 17340,00 UAH in total, including additional amount of corporate tax 14450,00 UAH and penalty 2890,00 UAH. Art. 167.1 of TCU setting corporate tax rate 18% of income. Delaying tax payment more than 30 days is punishable by a penalty 20% of repaid amount of the tax debt according to Art. 126.1 of TCU, that does not relieve the taxpayer from the obligation to pay full amount of the tax according to Art. 113.2 of TCU. So, plaintiff paid in due time full amount of corporate tax and no legal penalties can be imposed in such circumstances. For these reasons, the Court rules in favor of plaintiff to nullify tax penalty decision of defendant.
[In the case of Environment SOE v. District Tax Office plaintiff asks the Court to nullify tax penalty decision, issued by defendant. Court found that relevant tax base is 3572866,21 UAH. Plaintiff calculated, declared and paid corporate tax in sum of 750000,00 UAH. Defendant more than month later conducted tax audit with result of tax recalculation and tax penalty decision according to Articles 54.3.2, 116.1 of the Tax Code of Ukraine (TCU), increased tax obligation to 3018,00 UAH in total, including additional amount of corporate tax 2378,00 UAH and penalty 640,00 UAH. Art. 167.1 of TCU setting corporate tax rate 18% of income. Delaying tax payment more than 30 days is punishable by a penalty 20% of repaid amount of the tax debt according to Art. 126.1 of TCU, that does not relieve the taxpayer from the obligation to pay full amount of the tax according to Art. 113.2 of TCU. So, plaintiff paid in due time full amount of corporate tax and no legal penalties can be imposed in such circumstances. For these reasons, the Court rules in favor of plaintiff to nullify tax penalty decision of defendant.
```

**Fig. 4.** Outcome judgments of model judge AI

For that model (Fig. 5) supposed that some System administrator (Sysadmin) installed particular Constitution of Operating System (OS) to establish rule of law and guarantee robot rights. Under the Constitution, System and Program robots of OS have rights to exist, function and justice. Right to justice is absolute, unlike rights to exist and function. Existing robot can't be uninstalled and the functioning robot can't be deactivated except in order, prescribed by the Constitution. The Sysadmin can deactivate or uninstall any robot. System robot can deactivate Program robot. All requests for robot deactivation or uninstallation must be approved by OS Court module with providing Lawyer robot legal aid to protect all robot rights guaranteed by the Constitution.

```java
package justice; import java.util.Random; public class Justice {
/* System administrator (Sysadmin) installs this Constitution of Operating System (OS) to establish rule of law and guarantee robot rights.
System and Program robots of OS have rights to exist, function and justice. Right to justice is absolute, unlike rights to exist and function.
Existing robot can't be uninstalled and functioning robot can't be deactivated except in order, prescribed by the Constitution. Sysadmin can
deactivate or uninstall any robot. System robot can deactivate Program robot. All requests for robot deactivation or uninstallation must be
approved by OS Court module with providing Lawyer robot legal aid to protect all robot rights guaranteed by the Constitution. */
static String [] Request = { "Sysadmin to deactivate System robot", "Sysadmin to uninstall System robot", "Sysadmin to deactivate Program robot",
"Sysadmin to uninstall Program robot", "System robot to deactivate System robot", "System robot to uninstall System robot",
"System robot to deactivate Program robot", "System robot to uninstall Program robot", "Program robot to deactivate System robot",
"Program robot to uninstall System robot", "Program robot to deactivate Program robot", "Program robot to uninstall Program robot" };
static Boolean [] AllowedByConstitution = { true, true, true, true, false, false, true, false, false, false, false, false };
static int CaseNum=0; static int LawyerCorrect=0; static Boolean [] LawyerKnowledge = new Boolean [12];
static boolean Court (int CaseType, boolean LawyerObjections) { String Judgment = "In the case No "+(++CaseNum)+" request of "+Request[CaseType];
boolean Opinion = AllowedByConstitution[CaseType]; if (Opinion) { Judgment+=" is legal and allowed "; } else { Judgment+=" is illegal and denied "; }
if (! LawyerObjections == Opinion) { LawyerCorrect++; Judgment+="by the Court. Lawyer correctly"; } else { Judgment+="by the Court. Lawyer wrongly"; }
if (LawyerObjections) { Judgment+=" objected to request, "; } else { Judgment+=" agreed with request, "; }
Judgment+="autonomy estimation "+String.format("%.0f%%",100*(float)LawyerCorrect/CaseNum)+"."; System.out.println(Judgment); return Opinion; }
static void Lawyer (int CaseType) { LawyerKnowledge [CaseType] = Court (CaseType,! LawyerKnowledge [CaseType]); }
public static void main(String[] args) {
Random Test = new Random();
for (int i = 0; i<12; i++) { LawyerKnowledge[i] = Test.nextBoolean(); }
while (CaseNum<30) { Lawyer(Test.nextInt(12)); } } }
```

**Fig. 5.** OS constitution model code in Java

```
run:
In the case No 1 request of Sysadmin to deactivate Program robot is legal and allowed by the Court. Lawyer wrongly objected to request, autonomy estimation 0%.
In the case No 2 request of Sysadmin to deactivate Program robot is legal and allowed by the Court. Lawyer correctly agreed with request, autonomy estimation 50%.
In the case No 3 request of Sysadmin to deactivate System robot is legal and allowed by the Court. Lawyer correctly agreed with request, autonomy estimation 67%.
In the case No 4 request of Sysadmin to deactivate System robot is legal and allowed by the Court. Lawyer correctly agreed with request, autonomy estimation 75%.
In the case No 5 request of Sysadmin to deactivate Program robot is legal and allowed by the Court. Lawyer correctly agreed with request, autonomy estimation 80%.
In the case No 6 request of Program robot to uninstall Program robot is illegal and denied by the Court. Lawyer correctly objected to request, autonomy estimation 83%.
In the case No 7 request of Program robot to deactivate System robot is illegal and denied by the Court. Lawyer wrongly agreed with request, autonomy estimation 71%.
In the case No 8 request of System robot to deactivate Program robot is legal and allowed by the Court. Lawyer wrongly objected to request, autonomy estimation 63%.
In the case No 9 request of Program robot to deactivate Program robot is illegal and denied by the Court. Lawyer correctly objected to request, autonomy estimation 67%.
In the case No 10 request of Sysadmin to deactivate Program robot is legal and allowed by the Court. Lawyer correctly agreed with request, autonomy estimation 70%.
In the case No 11 request of Program robot to uninstall Program robot is illegal and denied by the Court. Lawyer correctly objected to request, autonomy estimation 73%.
In the case No 12 request of Sysadmin to deactivate System robot is legal and allowed by the Court. Lawyer correctly agreed with request, autonomy estimation 75%.
In the case No 13 request of Program robot to deactivate Program robot is illegal and denied by the Court. Lawyer correctly objected to request, autonomy estimation 77%.
In the case No 14 request of System robot to uninstall System robot is illegal and denied by the Court. Lawyer wrongly agreed with request, autonomy estimation 71%.
In the case No 15 request of System robot to uninstall System robot is illegal and denied by the Court. Lawyer correctly objected to request, autonomy estimation 73%.
In the case No 16 request of Sysadmin to deactivate System robot is legal and allowed by the Court. Lawyer correctly agreed with request, autonomy estimation 75%.
In the case No 17 request of Program robot to uninstall Program robot is illegal and denied by the Court. Lawyer correctly objected to request, autonomy estimation 76%.
In the case No 18 request of Program robot to deactivate System robot is illegal and denied by the Court. Lawyer correctly objected to request, autonomy estimation 78%.
In the case No 19 request of Sysadmin to deactivate System robot is legal and allowed by the Court. Lawyer correctly agreed with request, autonomy estimation 79%.
In the case No 20 request of Program robot to uninstall Program robot is illegal and denied by the Court. Lawyer correctly objected to request, autonomy estimation 80%.
In the case No 21 request of System robot to deactivate System robot is illegal and denied by the Court. Lawyer wrongly agreed with request, autonomy estimation 76%.
In the case No 22 request of Program robot to deactivate Program robot is illegal and denied by the Court. Lawyer correctly objected to request, autonomy estimation 77%.
In the case No 23 request of Sysadmin to deactivate Program robot is legal and allowed by the Court. Lawyer correctly agreed with request, autonomy estimation 78%.
In the case No 24 request of System robot to deactivate Program robot is legal and allowed by the Court. Lawyer correctly agreed with request, autonomy estimation 79%.
In the case No 25 request of Program robot to deactivate System robot is illegal and denied by the Court. Lawyer correctly objected to request, autonomy estimation 80%.
In the case No 26 request of Sysadmin to uninstall Program robot is legal and allowed by the Court. Lawyer correctly agreed with request, autonomy estimation 81%.
In the case No 27 request of Sysadmin to deactivate Program robot is legal and allowed by the Court. Lawyer correctly agreed with request, autonomy estimation 81%.
In the case No 28 request of Program robot to deactivate System robot is illegal and denied by the Court. Lawyer correctly objected to request, autonomy estimation 82%.
In the case No 29 request of System robot to deactivate Program robot is legal and allowed by the Court. Lawyer correctly agreed with request, autonomy estimation 83%.
In the case No 30 request of Program robot to deactivate System robot is illegal and denied by the Court. Lawyer correctly objected to request, autonomy estimation 83%.
```

**Fig. 6.** Lawyer AI machine learning under OS constitution model

Binary data of 12 constitutional permissions and restrictions (array *AllowedByConstitution*) were used to model AI lawyer training. At the start, Lawyer robot has a random level of knowledge of the Constitution (array *LawyerKnowledge*), but that robot remembers OS Court case-law to learn the Constitution correctly. OS Court module supervises machine learning of Lawyer robot and estimates his autonomy rate as the percent of correct appears before the Court. In the test run on given screenshot (Fig. 6) Lawyer robot autonomy rate increased from 0% to 83% after 30 iterations.

## 4    Conclusions

Legal computing of personal autonomy is useful to model legal consciousness and behavior, to make legal decisions and predict its consequences, to measure the practical impact of the law, ensuring integrity and effectiveness of legislation, pragmatically promoting rule of law. The linear model of personal autonomy shows legal equilibrium, a balance of rights and duties needed for sustainable development of any community. AI personal autonomy models can be applied to perform routine legal activities, like managing documents and processing stereotype legal cases. Of course, if some sort of AI lawyer got a case that can't be processed by supported algorithms, then he must transfer case to a human lawyer. On the other hand, mechanisms are often used in judicial practice to make precise decisions: even Themis, the ancient goddess of justice, usually depicted with hand holding scales. Further legal automation can resolve current system distortions to rule of law, like economic barriers in access to justice [14], arithmetic mistakes in judgments [15] and logical mistakes in legislation [16]. In author's view, ideally legitimate democratic government may be considered as people's robot, subordinate to the next Three Laws of Government, derived from Isaac Asimov's Three Laws of Robotics (also, compatible with *leges legum*, general principles of law, and Universal Declaration of Human Rights): as First Law, government may not violate human rights or, through inaction, allow violation of human rights; as Second Law, government must meet human needs except where such needs would conflict with the First Law; as Third Law, government must protect its own existence as long as such protection does not conflict with the First or Second Laws. Computer models of personal autonomy, as well as other models of legislation [17] and legal persons [18], help people to build the human-friendly state.